\documentclass[a4paper,fleqn]{cas-dc}

\usepackage{soul}
\usepackage{amsmath}
\usepackage[numbers,sort&compress]{natbib}

\usepackage{xcolor}
\definecolor{RED}{RGB}{156,78,90}

\def\tsc#1{\csdef{#1}∞{\textsc{\lowercase{#1}}\xspace}}
\tsc{WGM}
\tsc{QE}
\tsc{EP}
\tsc{PMS}
\tsc{BEC}
\tsc{DE}

\begin{document}

\begin{sloppypar}

\let\WriteBookmarks\relax
\def\floatpagepagefraction{1}
\def\textpagefraction{.001}
\shorttitle{Thermal response of the multiscale nanodomains}
\shortauthors{Mingbo Li et~al.}

\title [mode = title]{Universal thermal response of the multiscale nanodomains formed in $\it{trans}$-anethol/ethanol/water surfactant-free microemulsion}                      

\author[1]{Mingbo Li}[orcid=0000-0001-7891-3326]
\cormark[1]
\credit{Conceptualization, Methodology, Software, Validation, Formal analysis, Investigation, Writing original draft, Writing-review $\&$ editing, Visualization}
\address[1]{Center for Combustion Energy, Key laboratory for Thermal Science and Power Engineering of Ministry of Education, Department of Energy and Power Engineering, Tsinghua University, Beijing 100084, China}
\ead{mingboli@mail.tsinghua.edu.cn}

\author[1]{Yuki Wakata}
\credit{Methodology, Writing - Original Draft, Visualization, Validation}

\author[1]{Hao Zeng}
\credit{Writing - Original Draft, Visualization, Validation}

\author[1,2]{Chao Sun}
\cormark[1]
\credit{Conceptualization, Supervision, Writing - Review \& Editing, Funding acquisition, Project administration}
\address[2]{Department of Engineering Mechanics, School of Aerospace Engineering, Tsinghua University, Beijing 100084, China}
\ead{chaosun@tsinghua.edu.cn}
\cortext[cor1]{Corresponding author}

\begin{abstract}
\hl{\textbf{Hypothesis:}\\
Surfactant-free microemulsion (SFME), an emerging phenomenology that occurs in the monophasic zone of a broad category of ternary mixtures 'hydrophobe/hydrotrope/water', has attracted extensive interests due to their unique physicochemical properties. The potential of this kind of ternary fluid for solubilization and drug delivery make them promising candidates in many industrial scenarios.\\
\textbf{Experiments:}\\
Here the thermodynamic behavior of these multiscale nanodomains formed in the ternary $\it{trans}$-anethol/ethanol/water system over a wide range of temperatures is explored. The macroscopic physical properties of the ternary solutions are characterized, with revealing the temperature dependence of refractive index and dynamic viscosity.\\ 
\textbf{Findings:}\\
With increasing temperature, the ternary system shows extended areas in the monophasic zone. We demonstrate that the phase behavior and the multiscale nanodomains formed in the monophasic zone can be precisely and reversibly tuned by altering the temperature. Increasing temperature can destroy the stability of the multiscale nanodomains in equilibrium, with an exponential decay in the scattering light intensity. Nevertheless, molecular-scale aggregates and mesoscopic droplets exhibit significantly different response behaviors to temperature stimuli. The temperature-sensitive nature of the ternary SFME system provides a crucial step forward exploring and industrializing its stability.} 
\end{abstract}


\maketitle

\section{Introduction}

Microemulsion, a thermodynamically stable and optically isotropic transparent dispersion, shows extensive applications in numerous fields~\cite{bouchemal2004nano, srivastava2006detergency, anton2008design, anton2009universality, xu2017synthesis, liu2018co2, yan2021nanoprecipitation} since Hoar and Schulman introduced it~\cite{hoar1943transparent}. It can be formed in ternary or quaternary mixtures containing two immiscible fluids (a polar component, traditionally referred to as the water phase, and an apolar component, traditionally referred to as the oil phase) and at least one type of surfactant or cosurfactant. The presence of surfactant or cosurfactant in the interface between polar and apolar phases not only reduces interfacial tension but also gives flexibility,  making the surface more prone to bending, thereby facilitating energetically dispersion. There are three types of microemulsion structures, namely, oil-in-water (O/W), bicontinuous (BC), and water-in-oil (W/O), that can be clearly identified and distinguished in the composition diagram~\cite{moulik1998structure}. 

However, studies in recent years from different groups have demonstrated that the well-defined nano-domains with similar features as microemulsions also can been found in binary or ternary mixtures, even without the involvement of surfactants~\cite{subramanian2013mesoscale, schoettl2014emergence, hou2016surfactant, zemb2016explain, rak2018mesoscale}. The solution typically contains a special component called "hydrotrope". This is a type of amphiphilic molecule, but not surfactant molecules, that can significantly improve the solubility of insoluble compounds in aqueous solutions~\cite{hodgdon2007hydrotropic, xu2013surfactant}. The hydrotrope molecules usually consist of smaller non-polar tails and can not self-assemble to form either various micelles in bulk or ordered films at the water/oil interface~\cite{eastoe2011action}. Nevertheless they can constitute dynamic molecular clusterings involving hydrogen bonds with water molecules. They even have strong similarities with co-solvents at the nanoscale and macroscopic thermodynamic levels~\cite{kunz2016hydrotropes}. 

The spontaneously formed nano-structures are different from "standard" swollen micelles, and are given various names in different groups, for instance, detergentless microemulsions~\cite{smith1977oil}, "pre-Ouzo"~\cite{klossek2012structure}, surfactant-free microemulsions (SFME)~\cite{schoettl2014emergence, xu2018surfactant, yan2017modular, yan2021nanoprecipitation}, micellar-like structural fluctuations~\cite{novikov2017dual}, mesoscale solubilization~\cite{subramanian2013mesoscale, robertson2016mesoscale}, and ultraflexible microemulsions (UFMEs)~\cite{prevost2016small}. Thereafter, we use the term 'SFME' to represent this kind of nano-ordering. Moreover, the presence of thses mesoscale structures, ranging from several nanometers to hundreds of nanometers, have been confirmed as a general feature in water/hydrotrope mixtures containing hydrophobic compounds~\cite{diat2013octanol, xu2013surfactant, subramanian2014mesoscale, rak2018mesoscale}. The occurrence of SFME formation can be generalized: it is observed in a system containing two kinds of solvents (one polar and one apolar) that are basically immiscible, and a third one, hydrotrope, that is completely or at least partially miscible with both.

These well-organized nano-domains have prompted vigorous debates as to their origin and thermodynamic stability. They are indeed well-shaped entities and discrete domains that can exist for a long time from hours to years, rather than simple concentration fluctuations on the scale of a few nanometers. Conductivity and UV-Vis spectrum experiments help identifying three kinds of microstructures for SFME (resembling “ordinary” microemulsions) in the single-phase zone of the ternary phase system, including aggregates, bicontinuous phases, and reverse-aggregate structures~\cite{hou2016surfactant, han2022formation}. The characteristic scattering patterns corresponding to these three structures in the light scattering spectrum can be generally recognized. Similar observations were supported by direct optical observations by means of freeze-fracture and cryoscopic transmission electron microscopy~\cite{xu2013surfactant}. Nevertheless, those studies only present a macroscopic explanation of the observations, and an understanding on structuring from a microscopic perspective has been lacking until now. The ternary formulation 'octanol/ethanol/water' has been systematically investigated in the pioneering works~\cite{zemb2016explain} of Zemb et al. Two different nanoscale pseudo-phases, one octanol-rich and one water-rich, can be identified in the single-phase area near the miscibility gap by combining a manifold of scattering experiments and MD simulations~\cite{diat2013octanol, zemb2016explain}. The size of the aggregate is deduced to be on the order of $\sim$2 nm, and more generally, it has become clear that the aggregates detected in this class of ternary systems usually have a preferred size and interfaces, just like the classical microemulsions~\cite{diat2013octanol, schoettl2014emergence, schottl2016aggregation}. The ethanol, classified as a hydrotrope, can be found in both polar and nonpolar phases, and specifically is a major component of the interfacial film~\cite{schoettl2014emergence}. Its behavior resembles more the weak specific attraction of ions to interfaces~\cite{jungwirth2006specific}. A theoretical framework, that considers the balance between hydration force and entropy, has been proposed and suggested to be a basic principle for the spontaneous formation and stability of the nanostructures~\cite{zemb2016explain}. MD simulation results indicate that ethanol molecules are preferentially enriched near the interface and further promote hydration interactions~\cite{marcelja1997hydration, marvcelja2011hydration, donaldson2015developing, lopian2016morphologies}. The hydration interactions can effectively overcome entropy, resulting in the insertion of net repulsive forces between adjacent aggregates. From other perspectives, it has been proposed that the mesoscale ordering of these weak aggregates can be attributed to the attractive interactions at the molecular level~\cite{qiao2015molecular} or
competing short-range and long-range interactions~\cite{sweatman2014cluster, sweatman2019giant}. The insight into a atomic-level structure of the spontaneously formed nano-domains can only be obtained by combining molecular dynamics simulation and experiment.     

As an emerging field, progress is underneath in the comprehensive characterization of the structure and energetics of the weakly bound aggregates. In terms of application, one should not only understand the stability of the aggregates, but also master methods to destabilize them for various application purposes. For example, there have been a variety of molecular self-assembly pathways in nucleating solutions, highlighting the role of pre-nucleation microstructures (molecular or mesoscale scale clusters) in nucleation for a wide range of organic and inorganic systems~\cite{gebauer2008stable, davey2013nucleation, jawor2015effect}. Their stability and composition should be extremely sensitive to a wide variety of experimental scenarios, such as the physicochemical properties of the solutions or the surrounding environments. Further, the solution temperature, as a global factor, plays a crucial role in the formation and stability of SFME system. Revealing the dependence of SFME system on temperature may serve as a guide to the mind and help to answer the following questions: what is the phase behaviour of such liquid phases in which the multi-scale nano-domain coexists; what are the driving force behind their formation and stability; what is the emerging interfacial film.     

From our recent work~\cite{li2022spontaneously}, it is known that, multiscale nanostructures, whose correlation lengths exceed dimensions of individual molecules, can coexist in the single-phase area of ternary $\it{trans}$-anethol/ethanol/water solution. Both the bulk mixture and the nanoscale pseudo-phases present in it exhibit good stability with an extremely low Ostwald ripening rate under room temperature and atmospheric pressure. In this work, we expand our research on this topic, and embark on an systematic experimental work in an attempt to understand the thermodynamic stability of the SFME system throughout the monophasic region. We not only pay attention to the thermal response of multiscale nanostructurings in bulk, but also reveal the macroscopic physicochemical properties of ternary fluid over a wide temperature range. To our knowledge, this is the first report on how temperature affects the properties of ternary fluids, e.g., viscosity and refractive index.      

\section{Experimental Section}

\subsection{Chemicals and Solutions}

The ternary liquid mixture consisted of three components: water, ethanol and $\it{trans}$-anethol, namely, the composition of classical Ouzo system~\cite{lohse2020physicochemical}. Ultrapure water with initial pH 6.5 and a conductivity of 18.2 $\rm{M\Omega\cdot{cm}}$ was used throughout all the experiments and prepared with a Milli-Q purification system (Merck, Germany). Ethanol ($ 99.9\%$) was obtained from Beijing J\&K Co. Ltd. (China). $\it{trans}$-Anethol ($>99.0\%$) with $M_w = 148.20$ was purchased from Sigma-Aldrich (Germany). The physicochemical properties of these three components at 25$^{\circ}$C, including density $\rho$, dynamic viscosity $\eta$ and refractive index (RI), are listed in Table~\ref{tbl}. In the experiments all chemicals were of analytical grade and used as received.                           

\begin{table}[h]
\small
  \caption{\ Physicochemical properties of three pure components at 25$^{\circ}$C.}
  \label{tbl}
  \begin{tabular*}{0.48\textwidth}{@{\extracolsep{\fill}}lllll}
    \hline
    \textbf{Component} & \textbf{Formula} & \textbf{$\rho$ ($\rm kg/m^3 $)} & $\eta$ ($ \rm mPa \cdot s $) & \textbf{RI}\\
    \hline
$\it trans$-Anethol & $\rm C_{10}H_{12}O$ & 988.3 & 2.340 & 1.5610 \\
Ethanol & $\rm C_2H_6O$ & 785.46 & 1.057 & 1.3601 \\
Water & $\rm H_2O$ & 997.05 & 0.879 & 1.3325\\
    \hline
  \end{tabular*}
\end{table}

\subsection{Ternary Phase Diagram at Different Temperatures}

The ternary phase diagram of $\it{trans}$-anethol/ethanol/water system was constructed by visual titration method. The three components are measured and mixed by volume within a laboratory scale. The composition is defined based on the ratio of these three components to the total volume: $\phi_o$ ($\it{trans}$-anethol), $\phi_e$ (ethanol) and $\phi_w$ (water), where $\phi_o + \phi_e + \phi_w = 100\%$. The composition of the ternary mixture in the following is abbreviated as "$\rm O \phi_o E \phi_e W \phi_w$". We used the following steps to perform the experiment to outline the miscibility gap. A series of binary mixtures with different volume fractions of $\it{trans}$-anethol/ethanol (5:95, 10:90, 15:85, 20:80, 25:75, 30:70, 40:60, 50:50, 60:40, 70:30, 80:20, 90:10) were prepared in glass vial at room temperature (25$^{\circ}$C). The glass vial containing 10 mL of this binary mixture was tightened and then immersed in a circulating-water bath (PolyScience, PP15R-40), which can control temperature precisely. The water was also placed in this circulating-water to keep it at the same temperature. The water was then drawn and quickly added to the binary solution in successive $\rm 10\ \mu L$ increments using a micro-pipettor (ThermoFisher Scientific) with a range of $\rm1 \sim 10\ \mu L$ and an accuracy of $\pm 3.5 \sim 1.0\%$ until the appearance of the mixture changed from transparency to milky turbidity. The corresponding volume fraction of every composition in the formulation was recorded. Compared to the initial solution volume of 10 ml, the change in volume proportion caused was less than $1\%$. According to the transition point obtained at varying volume fractions of $\it{trans}$-anethol to ethanol, the phase-separation boundary line at a certain temperature can be determined. The same procedure was repeated three times for each set of binary mixtures. For each experimental condition with a certain temperature and a certain volume fraction of $\it{trans}$-anethol/ethanol, the same initial binary $\it{trans}$-anethol/ethanol solution was used in the three replicates, which were done within an hour. 

In addition, please note that for the case with room temperature 25$^{\circ}$C, the determination of the transition criterion from monophasic region to Ouzo region takes into account the turbidity of the ternary solution, which is measured using a turbidity meter (WZS-188, Shanghai INESA Scientific Instrument, China). However, the turbidity meter used in the experiments can not control the temperature of the sample, only suitable for measuring the turbidity of the sample at room temperature, namely 25$^{\circ}$C. We have to keep the mixture in the circulating-water bath to maintain a certain temperature. Nevertheless, we try to ensure the consistency and accuracy of the determination of the transition point through two steps. First, to slowly approach the phase separation line, only $\rm 10\ \mu L$ of water are added to the mixture at a time. Second, the Ouzo solution with a turbidity of 10.5 NTU at room temperature 25$^{\circ}$C is treated as a reference. We compared the appearance of the ternary solution obtained at the other temperatures with that of the Ouzo solution at 25$^{\circ}$C (turbidity: $\sim$10.5 NTU). When the two are identical or similar in appearance (the difference is almost indistinguishable to the naked eye), the transition is considered to have been reached, and the corresponding composition is considered the transition point. Here we have to say that there is an error in this qualitative comparison, but it is acceptable for this experiment.

\begin{figure}[!htb]
\centering
\includegraphics[scale = 0.45]{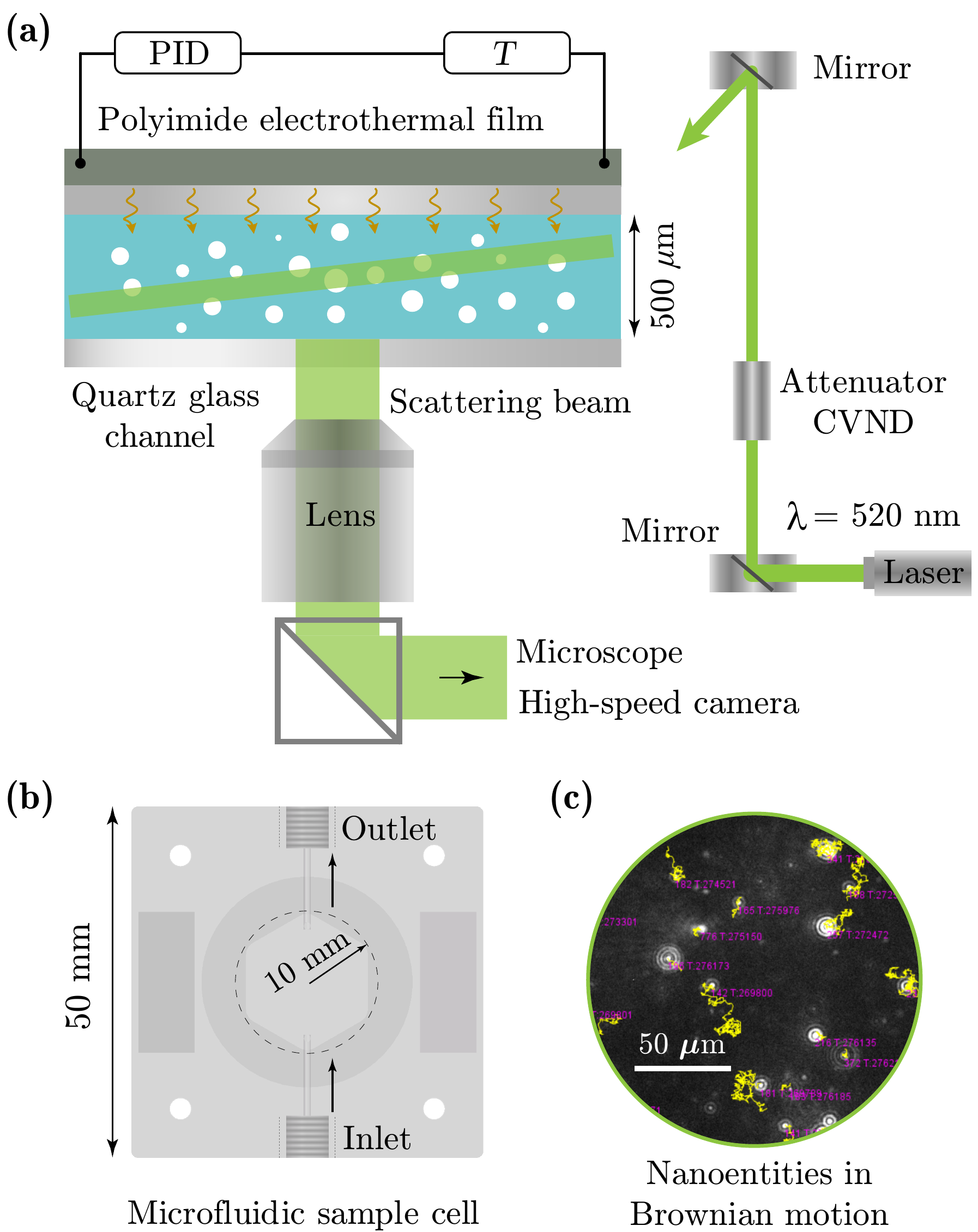}
\caption{(a) Schematic illustration of the NTA system (not-to-scale) used for the 'direct' imaging of mesoscopic droplets. (b) Top view of the microfluidic sample cell. (c) Trajectories of the nanoentities doing Brownian motion in the field of vision.}
\label{FIG1}
\end{figure}

\subsection{Dynamic Light Scattering Measurements (DLS)}

The DLS system (ZEN3700 Zetasizer NanoZSE, Malvern, UK) at a fixed scattering angle of $173^\circ$ was used to characterize the correlation function curves of the ternary mixtures and the corresponding size distribution of the nanoscale structures present in the bulk. A quartz sample cell with light path of 10 mm was used. For each measurement about 1 mL of the sample was loaded into the cell for each measurement. The cell was well sealed with a threaded lid during the measurement to reduce loss due to evaporation. The measurement was performed after the sample was equilibrated for 5 minutes at 25$^{\circ}$C. The nano-entities with hydrodynamic diameter ranging from 0.4 nm to 10 $\mu$m can be identified. 

The temperature range during DLS measurement is 10$^{\circ}$C $\sim$ 60$^{\circ}$C with an accuracy of $\leq\pm0.2^{\circ}$C ($\pm0.1^{\circ}$C at 25$^{\circ}$C). It is important that here we were restricted by the boiling temperature of volatile component, i.e., ethanol (78.3$^{\circ}$C). During the experiment, the ternary solution temperature was first reduced to 10$^{\circ}$C. We then progressively heated the sample from 10$^{\circ}$C to 60$^{\circ}$C. After setting a certain temperature in the software, the system took approximately $1\sim2$ minutes to reach a stable value. We then waited for 5 minutes until the system stabilized, meaning that the temperature of the solution was uniformly distributed, before proceeding with the measurement. We conducted two individual measurements for each sample, with each test repeated at least four times to obtain the average for better accuracy. Scattering intensities were measured by photon counting. The collected data were post-processed by considering the the transmittance of the laser after passing through the attenuator to obtain the total scattering intensity rate of the nano-entities (TSIR, kcps). 
 
\subsection{Nanoparticle Tracking Analysis Measurements (NTA)}

We designed and built a novel nanoparticle-tracking-analysis (NTA) apparatus to study the properties of the multiscale nanodomains during the thermal cycle. The concept of NTA, or named as ultramicroscope, and its application to the field of colloid chemistry is credited to Zsigmondy~\cite{siedentopf1902uber}. The device is mainly composed of three modules: a laser system, a detachable microfluidic sample cell and an imaging system with an optical darkfield microscope and a high-speed camera coupled, as shown in Figure~\ref{FIG1}(a). The sample cell is made of two pieces of quartz glass and a fluoroelastomer spacer. The lower plate (cross section $50\times50\ \rm{mm^2}$) is only used to hold and enclose the sample space. The upper cover plate is thicker and specially designed so that the laser beam (after multiple refractions) and solution can enter the sample space from opposite sides. Before passing through the sample the well-focused laser beam (power 150 mW, wavelength $\lambda = 520$ nm) is reflected twice by the mirror at different positions and then refracted twice by the prism-edged optical flat. The top view of this microfluidic sample cell is shown in Figure~\ref{FIG1}(b). The height of the hexagonal-like sample chamber is 500 $\mu\rm{m}$ and its maximum diameter is 20 mm. In the imaging module, the sufficient magnification is achieved by using a long working distance objective with 20X installed on an inverted optical microscope (IX73, Olympus, Japan). The upper plate is covered with a detachable polyimide electrothermal film, which is equipped with a PID control to adjust the temperature of the sample during measurement. Since there is no embedded refrigeration module, the temperature-controlling module is suitable for measurement from room temperature to 70$^{\circ}$C. 

The video sequences were acquired by combining the optical microscopy and a high-speed CCD camrea (xiD, XIMEA, Germany) operating at a framerate 30$\sim$40 fps. The optical resolution can reach about 227 nm/px. For each case, four videos lasting at least 60 s were recorded. After the measurements, the captured videos were evaluated via ImageJ/Fiji software~\cite{schindelin2012fiji}, specifically the plug-in NanoTrackJ ~\cite{wagner2014dark}. We can finally obtain the  number concentration of Brownian particles (see Figure~\ref{FIG1}(c)) and their size distribution, respectively. 


\subsection{Viscosity Measurements} 

The rheometer (Discovery Hybrid Rheometer, TA Instruments, USA) was used to measure the dynamic viscosity of the ternary solutions at varying temperatures (10$^{\circ}$C $\sim$ 60$^{\circ}$C). The temperature was varied in steps of 5$^{\circ}$C in the measuring range. During the measurements, a double gap cylinder (Couette) configuration was used. The viscosity of the sample was obtained at a constant shear rate. For a given sample, five individual measurements were performed. It is important to note that the measuring time of each run should be as short as possible such that the ethanol component would not be excessively lost due to its high volatility.

\subsection{Refractive Index (RI) Measurements}

The refractive index for the ternary solutions was obtained with an automatic refractometer (GR30, Shanghai Zhuoguang Instrument Technology, China) at various temperatures, ranging from 10$^{\circ}$C to 60$^{\circ}$C. Prior to taking measurements, we calibrated the refractometer at $25 ^{\circ}$C using deionized water according to the instrument's instructions. The measurement is made after the sample has reached the pre-set temperature of the refractometer. Particular attention should be paid to the duration of the measurement at high temperatures to prevent excessive evaporation of ethanol. For a given sample, three individual measurements were performed.

\section{Results and discussion}
 
\subsection{Ternary phase diagram}

\begin{figure}[!t]
\centering
\includegraphics[scale = 1.0]{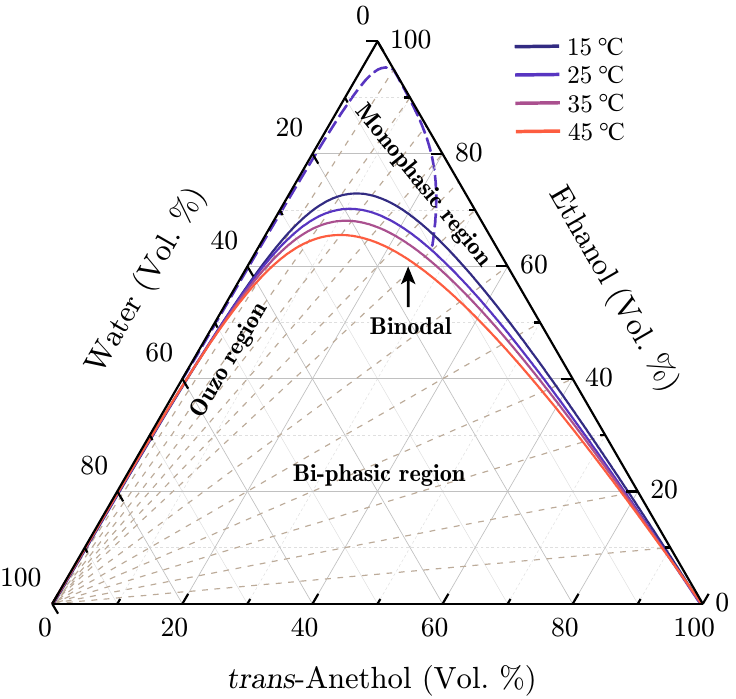}
\caption{Ternary phase diagram obtained at various temperatures. The brown dashed lines show paths of some composition coordinates from the titration experiments. The boundary lines of the three sub-regimes of the monophasic region at 25$^{\circ}$C are demonstrated (blue dashed lines).}
\label{FIG2}
\end{figure}

Ethanol can be miscible with water and $\it{trans}$-anethol in any proportion, but $\it{trans}$-anethol is insoluble with water. The $\it{trans}$-anethol/ethanol/water three-component mixture can outline a rich phase behaviors and further a ternary phase diagram, as reported in our recent study~\cite{li2022spontaneously}. The pseudo-monophasic domain can be divided into three regimes according to the nanostructures present in the ternary mixture: molecular solution, SFME (bicontinuous) and reverse aggregates. In this study, we focus on the miscibility limit, which marks the transition from macroscopic transparent pseudo-monophasic solution to milky Ouzo emulsion. The phase behavior of the $\it{trans}$-anethol/ethanol/water mixture was established at various temperatures (15$^{\circ}$C$\sim$45$^{\circ}$C), as depicted in Figure~\ref{FIG2}. The content of each component is expressed by volume fraction. In this ternary phase diagram, two distinct areas, namely monophasic and Bi-phasic regions, can be clearly identified. The boundary line (binodal) is obtained by fitting the measured experimental data (for more details please see Fig. S1 in Supplementary Material). The $\it{trans}$-anethol/ethanol/water SFME exhibits a sensitive temperature response. As the temperature increases, the miscibility-limit line passes downward evenly, causing a noticeable expansion of the monophasic area. In contrast, it is found that the Ouzo emulsion in the water-rich corner, which exist in metastable state~\cite{lohse2015surface}, is insensitive to temperature. The binodal lines near this region remain collapsed together and are not significantly offset.

\subsection{Effect of temperature on refractive index and viscosity}

\begin{figure}[!htb]
\centering
\includegraphics[scale = 0.70]{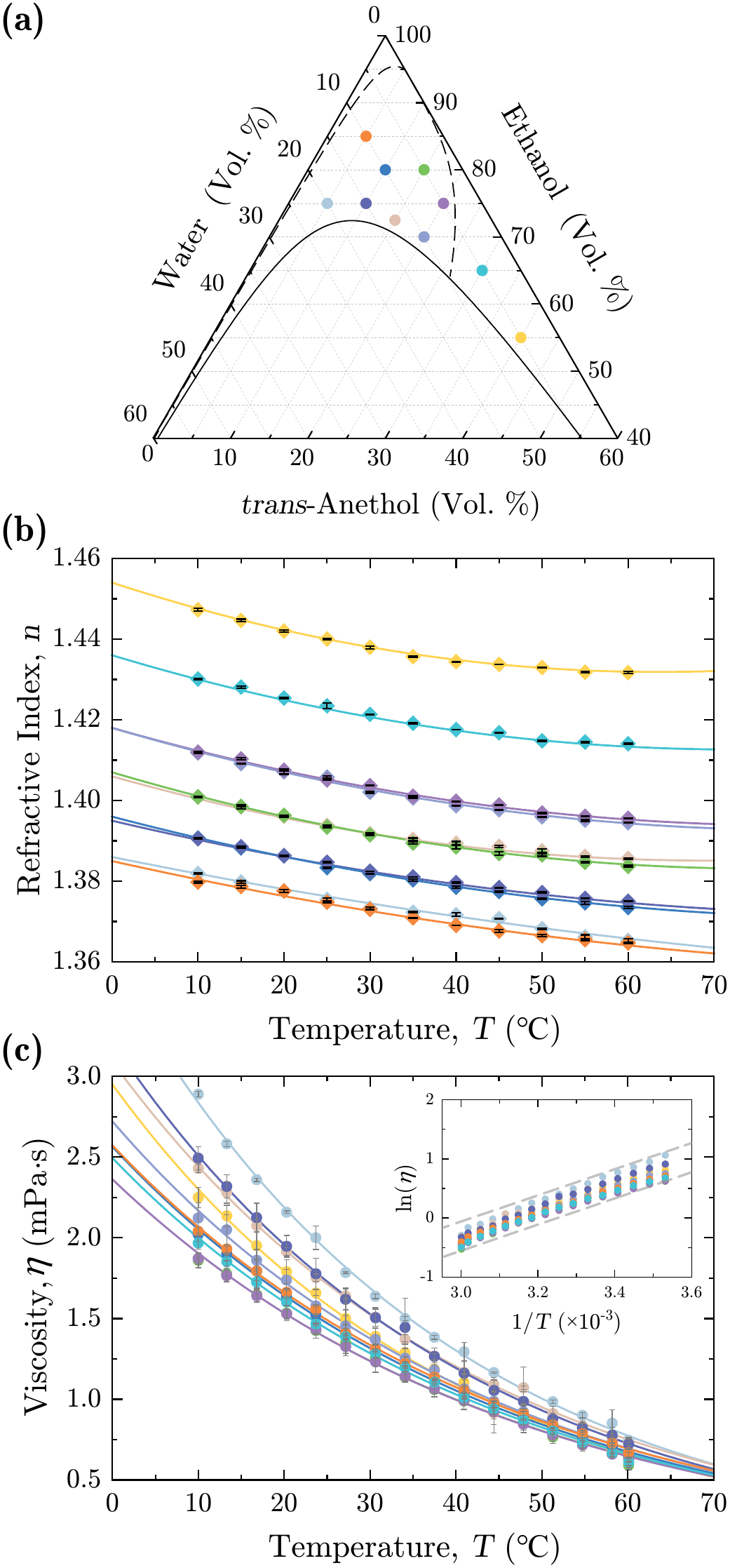}
\caption{(a) The symbols, marked in various colors, indicate different compositions of the ternary solutions measured. (b) Temperature dependence of refractive index of the ternary mixtures in monophasic region. (c) Temperature dependence of shear viscosity of the ternary mixtures in monophasic region. Error bars represent standard deviations from five independent measurements. The solid lines are best fits. Inset demonstrates the temperature dependence of viscosity fitted with VTF model.}
\label{FIG3}
\end{figure}

According to our previous work~\cite{li2022spontaneously}, the refractive index $n$ and dynamic viscosity $\eta$ of the ternary solution exhibit component-sensitive nature. The refractive index and shear viscosity data were collected from refractometer and rheometer measurements, respectively. Here samples with different component contents in the monophasic region were prepared at 25$^{\circ}$C, as shown in Figure~\ref{FIG3}(a). Then the temperature dependence of these two parameters is depicted. Figure~\ref{FIG3}(b) shows how temperature affects the shear viscosity $n$ for the samples marked in Figure~\ref{FIG3}(a). The $n$ decreases gradually with increasing temperature. At the same temperature, the $n$ depends only on the volume fraction of $\it{trans}$-anethol phase. This is understandable, the refractive indices of ethanol and water are similar, and both are lower than that of $\it{trans}$-anethol.   

The temperature dependence of the dynamic viscosity $\eta$ is presented in Figure~\ref{FIG3}(c). It can be found that the viscosity decreases obviously when the temperature increases, especially in the low temperature range. In high-temperature range, for instance, above 60$^{\circ}$C, interestingly, the viscosity of all samples collapses to nearly the same value at a given temperature. There are many models that have been proposed to describe the viscosity-temperature relationship. Among them, the two most widely used models are the Arrhenius equation: $\eta = A{\rm{exp}}(\varepsilon/RT)$ and the Vogel-Fulcher-Tamman (VFT) equation: $\eta = A{\rm{exp}}[{DT_0}/{(T-T_0)}]$, where $A$ is correlation parameter, $\varepsilon$ is the activation energy for viscous flow, $R$ is the gas constant, $T$ is the Kelvin temperature, $D$ is constant which is inversely proportional to the fragility of the liquid, and $T_0$ is the temperature at which the configurational entropy content of the supercooled liquid would vanish in a cooling process of infinite time scale. However, most models were extensively studied and reported for pure components but had limited application for mixture~\cite{kerschbaum2004measurement, restolho2009viscosity, yuan2009predicting}. 

We found that, after fitting, the viscosity measured in our experiments follow a similar exponential behaviour with rising temperature, as shown in the inset of Figure~\ref{FIG3}(c). The exponential behaviour can be described with the relationship ${\rm ln(\eta)} = A/{\it{T}} - B$. The values of $A$ and $B$ are obtained with a correlation. Our results show that the previously developed empirical models can still be applied to describe the experimental value of viscosity versus temperature for ternary mixtures. These findings indicate that multicomponent solutions with intermolecular solubility (binary or ternary) may follow a generic nature when considering viscosity. The presence of nanoscale pseudo-phases is not sufficient to affect some of the macroscopic physical properties of the multicomponent solution.

\subsection{Effect of temperature on SFME nano-domains}

\begin{figure}[!htb]
\centering
\includegraphics[scale = 0.7]{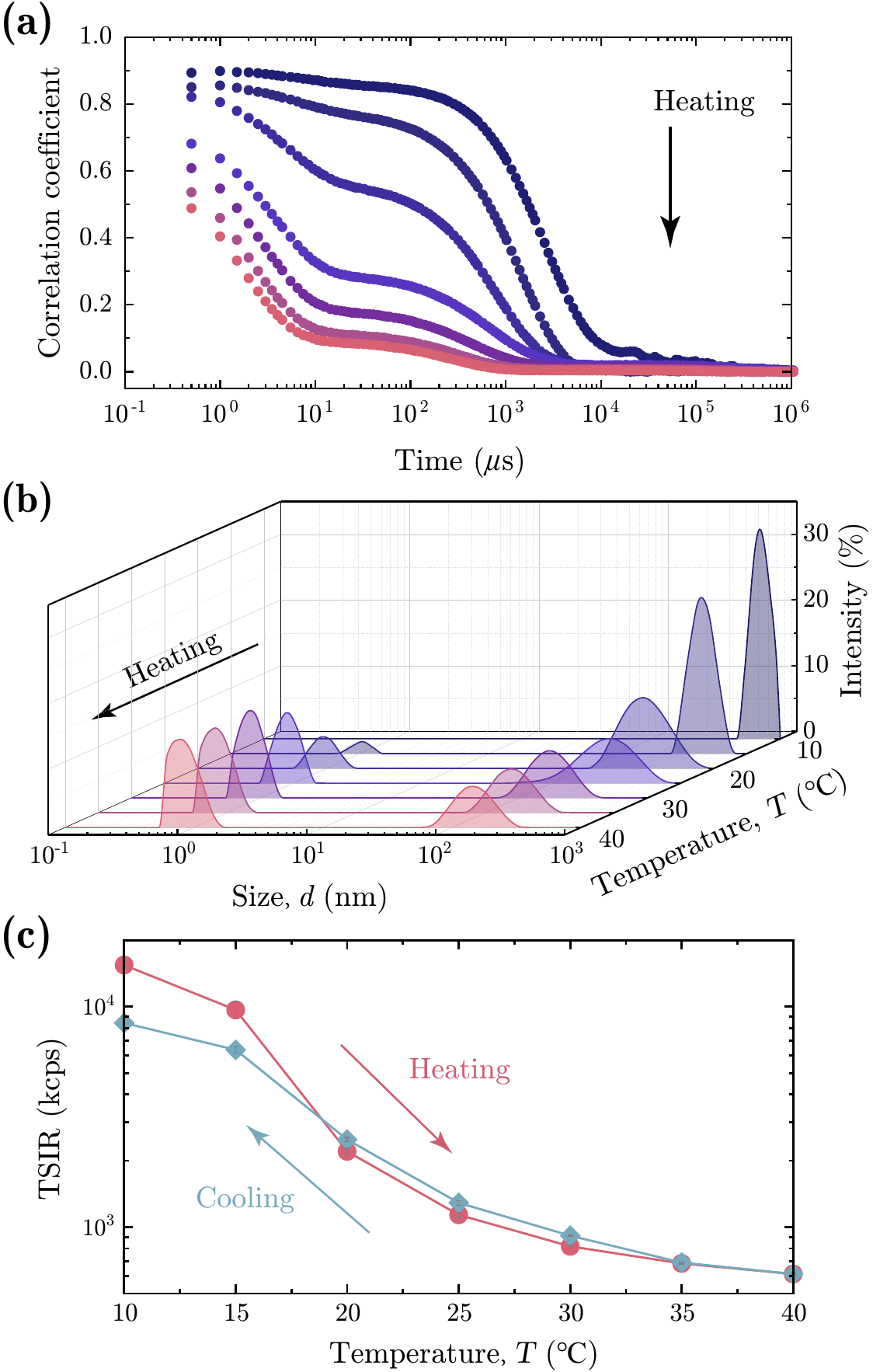}
\caption{Reversible temperature response for the SFME nano-domains formed in the monophasic mixture with compositions $\it trans$-anethol 15$\%$/ethanol 72.5$\%$/water 12.5$\%$. (a) DLS intensity auto-correlation functions and (b) corresponding SFME nano-domain size distributions obtained at different temperatures during heating (from 10$^{\circ}$C to 40$^{\circ}$C). (c) Dependence of the total scattering intensity rate (count rate, kcps) on temperature during the thermal circle. Each experimental point represents the mean and standard deviation across three measurements.
}
\label{FIG4}
\end{figure}

\begin{figure}[!htb]
\centering
\includegraphics[scale = 0.95]{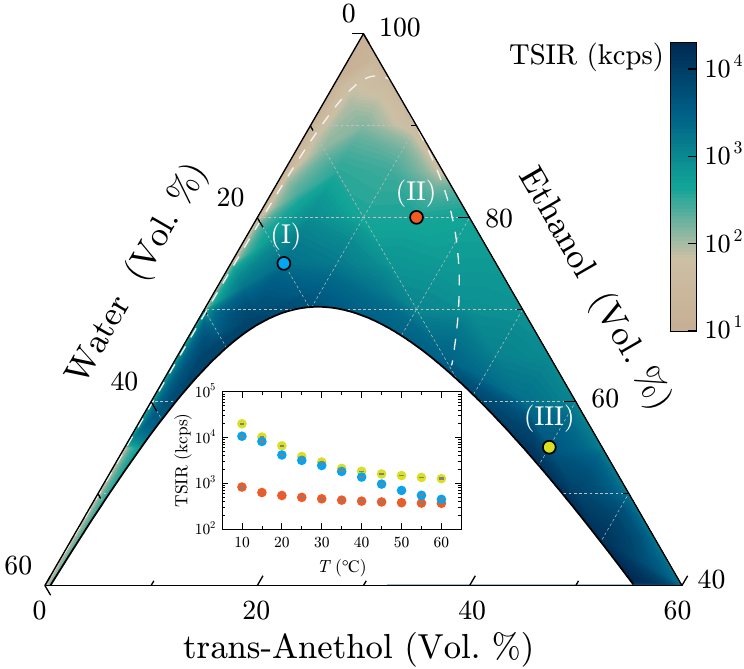}
\caption{Color-coded contour plot of the total scattering intensity rate for the entire monophasic region obtained at 25$^{\circ}$C. Three representative samples are marked. The inset presents the change of TSIR with the temperature increase for three cases. Each experimental point represents the mean and standard deviation across five measurements.}
\label{FIG5}
\end{figure}

\begin{figure}
\centering
\includegraphics[scale = 0.80]{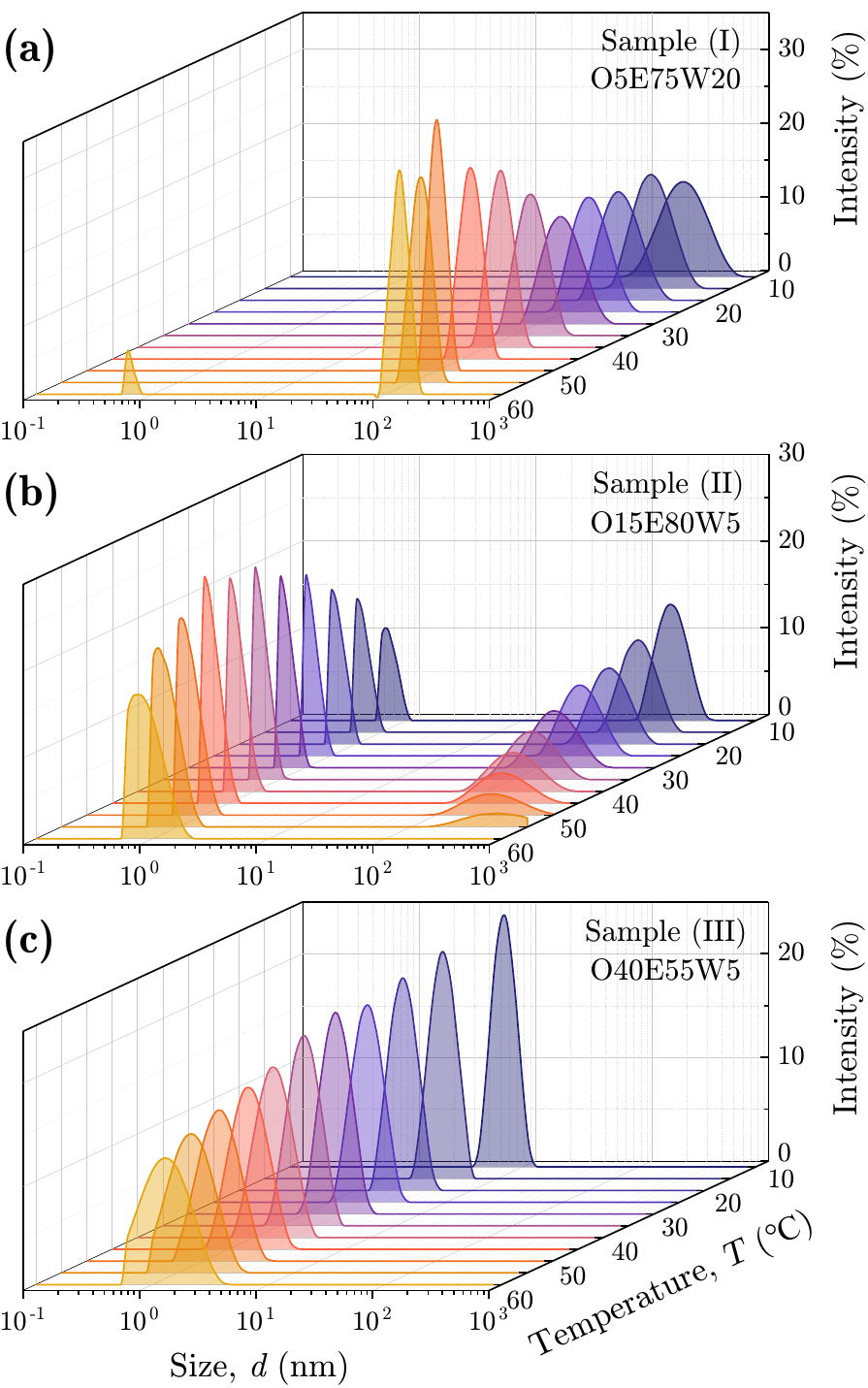}
\caption{Temperature dependence of the multiscale nano-domains formed in the monophasic region for the three cases marked in FIG. 5.}
\label{FIG6}
\end{figure}

We next investigate the temperature response of multiscale nano-domains coexisting in ternary solutions, that is, the monophasic region of the ternary phase diagram. In Figure~\ref{FIG4}, a typical case, where a sample undergoes a thermal circle, was shown. The preparation of the ternary system, consisting of 15$\%$ $\it trans$-anethol, 72.5$\%$/ ethanol and 12.5$\%$ water, was tested. As shown in Figure~\ref{FIG4}(a), a double exponential pattern, essentially revealing two different dynamic relaxation modes of Brownian particles in the suspension, can be observed.

The intensity auto-correlation functions obtained from in-situ measurements became less and less pronounced progressively when temperature increases from 10$^{\circ}$C to 40$^{\circ}$C. The spatial coherence factor decreases from 0.9 to about 0.5 monotonously. Here the two dynamic modes substantially suggest the presence of at least two classes of nanostructures with well-defined size in the ternary solution. With viscosity and RI corrections, the dimensions of these nanoscale structures at different temperatures were determined, as shown in Figure~\ref{FIG4}(b). A bimodal distribution is observed. It can be found that, in this case the multiscale nanodomains coexist, which spans two orders of magnitude in the hydrodynamic size, corresponding to the aforementioned two relaxation modes implied in the scattering-light signal. Since falling in the SFME region, these molecular-scale ($\sim$1 nm) and mesoscale ($\sim$100 nm) inhomogeneities are $\it trans$-anethol-rich aggregate and mesoscopic droplet, respectively, as named in our previous study~\cite{li2022spontaneously}. 

It is clear that, with increasing temperatures, the right peak, representing the mesoscopic droplet, decays sharply and shifts continuously to samller sizes. In contrast, for the molecular-scale aggregates, the size distribution broadens with a strong peak appearing. Since the effect of rising temperature on these nanostructures is so significant, what happens if the temperature drops instead? As observed in Fig. S2 (see Supplementary Material), interestingly, when the ternary solution's temperature cools back to initial temperature 10$^{\circ}$C, their size distribution eventually returns to the original state exactly. Evolution of the total scattering intensity rate (TSIR) of the multi-scale nanodomains in this thermal circle is shown in Figure~\ref{FIG4}(c). This parameter reflects the global information about scattering sources in a spatially uniform flow field. The scattering intensity rate decays exponentially with increasing temperature. After cooling, despite the deviation, it is nearly restored to the initial level. The deviation may be attributed to a small amount of ethanol condensing on the cell wall during cooling, which changes the composition of the sample. The present experimental findings clearly indicate that the nanodomains formed in ternary solution can be regulated precisely by changing the ambient temperature, demonstrating a superior reversibility. Therefore, unless otherwise specified below, all data relating to temperature are obtained when the sample is heated.

\begin{figure*}
\centering
\includegraphics[scale = 0.75]{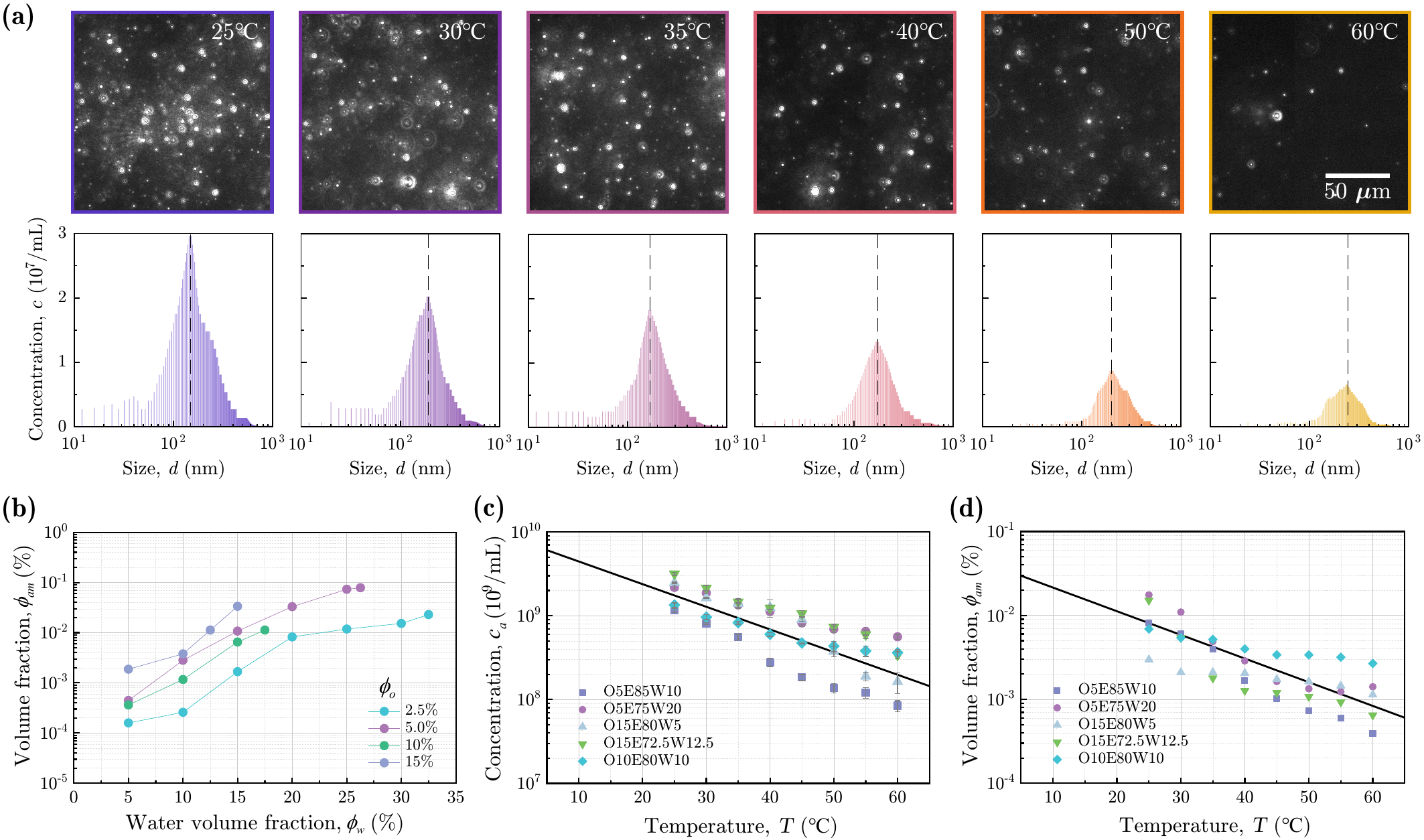}
\caption{(a) Evolution of mesoscopic droplets during heating from 25$^{\circ}$C to 60$^{\circ}$C in NTA experiments (composition of the ternary mixture: $\phi_o = 10\%$, $\phi_e = 80\%$, $\phi_w = 10\%$).  Upper panel: a series of experimental snapshots capturing the presence and dynamics of mesoscopic droplets under various temperatures. Lower panel: the corresponding size-concentration distribution histogram. (b) Volume fraction of the $\it{trans}$-anethol phase in the form of mesoscopic droplets in the bulk to the total added, $\phi_{md}$, for a ternary solution with different compositions. The ternary mixtures included in each set were obtained by fixing the volume fraction of $\it{trans}$-anethol and increasing the volume fraction of water. (c$\sim$d) Effect of temperature on the number density and the volume fraction of mesoscopic droplets for five cases.}
\label{FIG7}
\end{figure*}

Since temperature is intimately related to the TSIR, we measured systematically the scattering intensity rate of the nanostructures present in the entire single-phase region. The result is shown in Figure~\ref{FIG5}. One can see that, traversing the entire monophasic region, TSIR changes by at least three orders of magnitude. The scattering signal in the molecular solution is negligible. Approaching the phase separation boundary, TSIR is significantly enhanced, indicating a much prominent contribution from the nanostructurings, i.e., the mesoscopic droplets. Additionally, the TSIR in the reverse-aggregate region is more pronounced. Considering that the size of water-rich aggregates in this region is only around 5 nm, this means that its number density would be very high. 

We have further expanded the temperature range that the sample can withstand. When exposed to atmospheric pressure, the thermal response of multiscale nanodomains is explored by altering the solution temperature from 10$^{\circ}$C to 60$^{\circ}$C sequentially. Three typical samples are selected, as marked in Figure~\ref{FIG5}. The inset of Figure~\ref{FIG5} depicts the trend of TSIR versus temperature for the three cases. As expected, heating causes the TSIR to drop monotonically. The TSIR drop is more pronounced when the compositions of ternary solution are near the miscibility gap in the ternary system, reaching $\sim 90\%$. The reduction of TSIR may be related to the structural changes that occur at the molecular or mesoscopic scale. Correspondingly, the size distributions of the coexisting nanostructures measured at different temperatures for these three samples are shown in Figure~\ref{FIG6}(a$\sim$c). For multi-scale nanostructures formed in the SFME regime, i.e., Sample $\rm \uppercase\expandafter{\romannumeral1}$ and Sample $\rm \uppercase\expandafter{\romannumeral2}$, the mesoscopic droplets display weak stability at high temperatures, and either decrease in size (Figure~\ref{FIG6}(a)) or even disappear completely (Figure~\ref{FIG6}(b)). It is understandable that there is a competition between these two characteristic length-scales in terms of scattering intensity. As the temperature increases, the scattering from the $\it trans$-anethol-rich aggregates becomes comparable with or overwhelms that from the mesoscopic droplets. When heated to 60$^{\circ}$C, for Sample $\rm \uppercase\expandafter{\romannumeral1}$, a sharp peak appears at around 1 nm; for Sample $\rm \uppercase\expandafter{\romannumeral2}$, the aggregates in the bulk are so dominant that the size of mesoscopic droplets becomes undetectable. Here we emphasize that, when the solution temperature is less than 60$^{\circ}$C, the absence of peak of aggregates does not mean that there are no aggregates in the bulk solution. Instead, the scattering-light signals of these aggregates are easily masked by that of such large mesoscopic structures. Furthermore, the correlation length of $\it trans$-anethol-rich aggregates remains nearly constant throughout the heating process, demonstrating the superior stability. For water-rich aggregates with unimodal distribution (Sample $\rm \uppercase\expandafter{\romannumeral3}$), the sharp peak collapses and shifts to the left with increasing temperature, as shown in Figure~\ref{FIG6}(c). The mean size decreases from $\sim$7 to $\sim$2 nm, which is a collective behavior.    

NTA measurements can provide direct "visualizing" observation on the evolution of mesoscopic droplets in the whole thermal cycle. NTA system enables direct counting of nanoparticles and the analysis of tracks of individual objects in real time~\cite{rak2018mesoscale, hsu2022observation}. In Figure~\ref{FIG7}(a), a series of microscopic snapshots for a sample when heated from 25$^{\circ}$C to 60$^{\circ}$C are exhibited. Note here that the molecular-scale aggregates are too small to be visible in the view of NTA system (minimum resolution: 20 nm)~\cite{ma2022measurement}. The light intensity from the suspended particles within the focal volume drops dramatically upon heating, and the exposure time and gain intensity have to be adjusted synchronously during the imaging recording to obtain clear videos. The corresponding size-concentration distribution is shown at the bottom. A well-defined monomodal peak under a certain temperature, centered around 200 nm, is observed. It is clear that, with increasing temperature, most mesoscopic droplets gradually disappear in the view and their size distribution becomes increasingly narrower. Even at high temperatures, the number concentration of observed mesoscopic droplets in the scattering volume is quite low, making it impossible to statistically analyze their size distribution. This is a solid evidence that the precipitous decline of the TSIR is due to the disappearance of mesoscopic droplets. After cooling, the mesoscopic droplets nucleate accurately and appear in the view again. Nevertheless, the mean size of the mesoscopic droplet population shows weak dependency on the temperature.          

In SFME regime, the thermodynamic properties of mesoscopic droplet provide a perspective on how these nanostructurings destabilize at elevated temperatures. We wonder the proportion of the $\it trans$-anethol phase present in the form of mesoscopic droplets in the ternary solution. The volume fraction of $\it trans$-anethol in the form of mesoscopic droplet to the total volume mixed can be calculated roughly using the equation $\phi_{am} = {V_{md}N_{md}}/{\phi_a}$, where $V_{md}$ is the volume of single mesoscopic droplet, $N_{md}$ is the number concentration of mesoscopic droplets present in the bulk ternary solution, and ${\phi_a}$ is the volume fraction of $\it trans$-anethol initially added to the mixture. Here for each sample, the mean diameter is considered in the calculation. The volume fractions of $\it trans$-anethol in the form of mesoscopic droplets in different ternary solutions are shown in Figure~\ref{FIG7}(b). However, somewhat surprisingly, after mixing only trace amounts of the component of $\it trans$-anethol, typically $0.0001\%$<$\phi_{am}$<$0.1\%$, are incorporated into the mesoscopic droplet. More than $99.9\%$ of the populations exist in the form of molecules or molecular-scale clusterings, which cannot be resolved optically in the view. It also can be seen that $\phi_{am}$ increases exponentially with the increase of the volume fraction of added water $\phi_{w}$. The increase of $\it trans$-anethol composition is also conducive to the formation of mesoscopic droplets. Nevertheless, please note that such mesoscopic structures with a scale of $\sim$100 nm in size is not universal, and only were identified in some specific ternary systems~\cite{prevost2016small, robertson2016mesoscale, chiappisi2018looking}. In particular, mesoscopic droplets may also be observed only in a specific formulation range of a ternary system. The nature and origin of this type of mesoscopic structure are still open questions, and will be explored in future study. Even so, the light intensity scattered by such a small number of mesoscopic droplets dominates the whole.

Furthermore, we quantify the number concentration and volume fraction of mesoscopic droplets $\phi_{md}$ with increasing temperatures, as shown in Figure~\ref{FIG7}(c$\sim$d). As the temperature rises, a substantial decline of the concentration $c_{a}$ or volume fraction $\phi_{am}$ with a $\sim e^{-0.06T}$ dependence is observed. All sets of the results exhibit similar trends, demonstrating that the stability of mesoscopic structures weakly depends on the formulation of the ternary mixture. Collectively, our results demonstrate that, the decrease in the scattering light intensity in the ternary solution upon increasing temperature is due to the thermodynamic instability of the mesoscopic droplets, rather than the collective reduction in their size. It is highly likely that hundreds of homogeneous nucleation events would suddenly occur within the bulk solution upon cooling. 
        
\begin{figure}[!htb]
\centering
\includegraphics[scale = 0.98]{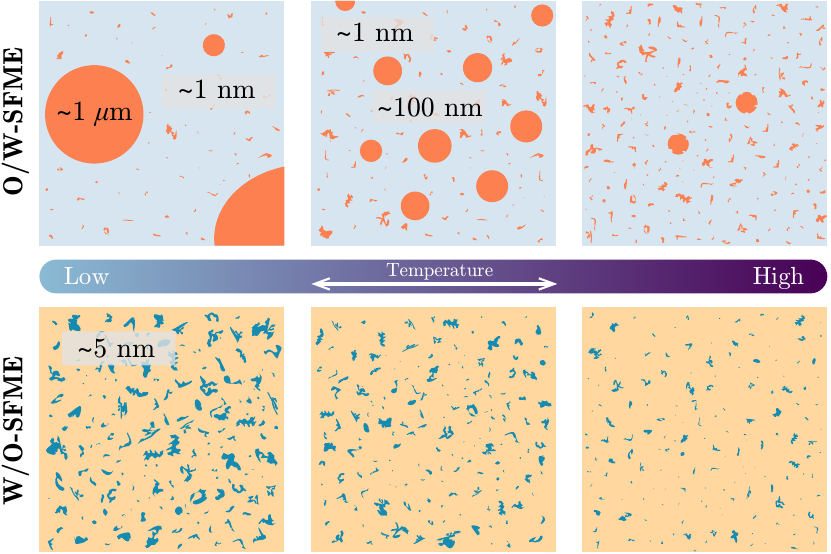}
\caption{Schematic representation illustrating the evolution of nano-domains when increasing temperatures (not to scale). The $\it{trans}$-anethol-rich and water-rich nanodomains are represented in orange and blue, respectively.}
\label{FIG8}
\end{figure}

In Figure~\ref{FIG8}, we summarize our findings and depict the representative snapshots, which capture how these multiscale nanodomains and phase behaviors reversibly evolve with temperature changes. Here we refer to previous pioneering works~\cite{zemb2016explain, schottl2019combined} on the morphology of direct and reverse molecular-scale aggregates. Macroscopically increasing temperature will drive the system into the SFME state from Ouzo state, where the Ouzo emulsions can coexist with the SFMEs~\cite{prevost2021spontaneous}. In SFME region, with increasing temperature, the mesoscopic droplets lose their thermostability and progressively vanish, while the molecular-scale aggregates show extreme stability with almost constant size. Within the temperature range of our study, these dispersed $\it{trans}$-anethol-rich aggregates have no tendency to lose stability and eventually transform into continuous phase. Moreover, we believe that the unstable mesoscopic droplets in the thermal field may be partially converted into aggregates instead of completely forming molecular solutions, as evidenced by the dramatically enhanced scattering-light intensity. Upon cooling, the disappeared mesoscopic droplets nucleate again spontaneously with right size, indicating that it is not a stochastic process. There would be no distinct onset or cut-off temperatures at which the nucleation of the mesoscale structures get start or ends. It is also remarkable that at ambient temperature the mesoscopic droplets can maintain relative stability for a long period of time, forming a nearly homogeneous and metastable colloidal system. The temperature-sensitive nature of mesoscopic droplets may emphasize that they are kinetically trapped structures, rather than thermodynamic equilibrium structures. In the reverse-SFME regime, as the temperature is raised, the water-rich aggregates only monotonously shrink; after the temperature recovers, they grows to the initial state.

Why do the multi-scale nanodomains formed in ternary $\it{trans}$-anethol/ethanol/water system exhibit such sensitive temperature-responsive behavior? The ternary samples open to the thermal environment are subjected to numerous factors. At first glance, there is a positive correlation between temperature and molecular solubility of $\it trans$-anethol. A certain nucleation barrier should be overcome to form mesoscopic droplet. With increasing temperature, the nucleation barrier is elevated, extending the miscibility limit in the ternary phase diagram. More importantly, the instability or collapse of the mesoscopic droplets undergoing heating may be due to the destruction of the protective shell, that surrounds the $\it trans$-anethol core. The molecular dynamics simulation has demonstrated that, the ethanol molecules, one class of amphipathic molecule that induces the interfacial energy, prefer to enrich at the interface of the mesoscopic droplets~\cite{schoettl2014emergence, hahn2019ab, han2022formation}. Moreover, the hydrogen-bonded shell especially formed between water and ethanol molecules can be multilayered and serve as an interfacial film~\cite{subramanian2013mesoscale}. The water molecules inside the film do not have a strong interaction with the hydrophobic molecules, namely, $\it trans$-anethol, aggregating in the core. This shell separates the core from the bulk phase of the solution to keep it stable. As the temperature increases, the hydrogen-bond interaction may be weaken, and hence the 'protective shell' of the mesoscopic droplets would be broken. The amphiphilic molecules may escape from the interface phase to the bulk, eventually resulting in the thermodynamically collapsing of the mesoscale structurings. Furthermore, a theoretical framework that takes into account the hydration force and entropy to determine the formation of microstructures and their correlation lengths has been established in the case of hydrotrope-based microemulsions~\cite{zemb2016explain}. Any increase or decrease in the characteristic size of these nanodomains consumes overall free energy. The temperature may also affect the structure and stability of the double layer depending on several factors, including the concentration of the electrolyte, the degree of ion hydration, and the electrical properties of the solution and the charged surface~\cite{marcelja1997hydration}. The hydration force, or a repulsive hydration interaction, formed by a high hydroxyl surface density, is very important near the interface, even in pure water. The Mar$\rm \check{c}$elja-Radi$\rm \acute{c}$ theory for hydration forces obtained in pure water has been well extended to multicomponent solvents~\cite{marvcelja2011hydration, zemb2016explain}. Specifically, at increasing temperatures, water molecules become less structured and less hydrogen-bonded, and the degree of ion hydration may also decrease, both of which can lead to a decrease in the thickness and stability of the electrical double layer. This may further break the hydration-entropy balance and in turn induce the instability of these multi-scale nanodomains. 

Besides, the thermal response of these multiscale nanodomains may open up a perspective to study their origin. The appearance of these tiny nuclei as a new phase within a practically homogeneous liquid in the vicinity of the miscibility limit may result from heterophase density fluctuations, which was first proposed and generalized by Frenkel in 1939~\cite{frenkel1939general}. Heterophase fluctuations refer to fluctuations in the relative amounts of different phases within a material. They can give rise to pre-transition phenomena, which are observable changes in the bulk properties prior to the actual phase transition. The critical point hence is no longer a "normal" or "specific" transition point about composition. For example, precrystallization phenomena can be observed in a liquid in the vicinity of its thermodynamic crystallization point due to the formation of small crystalline nuclei. The size and number of these domains are also temperature sensitive. In particular, the emergence of these nanostructures is unlimited and do not occur only in thermodynamically unstable states, such as supercooling or superheating. The transition seems to be a very slow process, which is difficult to define practically as a stable statistical equilibrium or unstable statistical equilibrium~\cite{gebauer2014pre}. Indeed, this is still an open question and a clear molecular view for the temperature response of these multiscale nanodomains is needed in our opinion.

\section{Conclusions and outlook} 

The surfactant-free microemulsion, an emerging phenomenology that occurs in a wide class of ternary systems 'hydrophobe/hydrotrope/water', encompasses numerous thermodynamic and structural behaviors that have yet to be revealed. The phase behavior and the nanostructurings at the nanometer scale (one aggregate with a characteristic size of the order of 1 nm and one mesoscopic droplet with a characteristic size of the order of 100 nm) have been clarified in our recent work~\cite{li2022spontaneously}. In the present work, we expanded on previous work and have demonstrated that these multiscale nanodomains formed in the ternary $\it{trans}$-anethol/ethanol/water system can be reversibly tuned by altering the temperature. The macroscopic bulk SFME solution and the microscopic nanodomains spontaneously formed are characterized and identified over a wide temperature range (10$^{\circ}$C $\sim$ 60$^{\circ}$C). 

The results revealed that, with increasing temperature the ternary system shows extended areas in the monophasic zone. The temperature dependence of the macroscopic physical properties of the ternary solutions, such as refractive index and dynamic viscosity, has been quantitatively characterized over the whole composition range of the monophasic region. Both refractive index and viscosity decrease monotonously with increasing temperature. Especially for the ternary mixture's viscosity, there is a catastrophic exponential decay that can be captured by the Vogel-Fulcher-Tamma equation. Our study extends the existing viscosity-temperature empirical model to multicomponent fluids~\cite{kerschbaum2004measurement, restolho2009viscosity, yuan2009predicting}, and demonstrates that multicomponent solutions with intermolecular solubility may follow a generic nature.


Moreover, increasing the temperature can destroy the long-term stability of the multiscale nanodomains in equilibrium, with an exponential decay in the scattering light intensity. This may promote the transformation of SFME solution into a real intermolecular-miscible solution, similar to demulsification. Upon cooling, both the phase behavior and the nanodomains return to the initial state exactly, showing an excellent reversibility. Combined with NTA measurements, we demonstrated that the attenuation of scattering light intensity at elevated temperature is primarily attributed to the disappearance of mesoscopic droplets, whose $\it{trans}$-anethol content accounts for only a tiny fraction (0.0001$\sim$0.1$\%$) in the bulk. Our study highlights the important role of such mesoscopic structures (characteristic size scale $\sim$100 nm) in thermal response, which are not universally recognized in ternary SFME systems~\cite{diat2013octanol, prevost2021spontaneous}. It is a new type of colloidal particle. Their number concentration decays dramatically with temperature in an exponential relationship, whereas their size, typically 100$\sim$300 nm in diameter, shows a weak temperature dependence. Further, the mesoscopic droplets can nucleate rapidly and spontaneously with accurate size and distribution once cooled. The temperature-sensitive nature of mesoscopic droplets may emphasize that they are in thermodynamic equilibrium state once formed, and clarify a way to study their origin. In contrast, the molecular-scale $\it{trans}$-anethol-rich aggregates in the thermal circle are remarkably stable. In the reverse-SFME, the water-rich aggregates, however, shrink continuously with increasing temperature, presenting a collective thermal response behavior. 

Another yet open question is the specific molecular structure of these multiscale nanodomains, which will help to understand their thermal-response behavior. We hope that this work will trigger more MD simulation works pushing the size and time scales limitations currently covered so as to rationalize the experimental results. Finally, our study provide a simple method to switch the phase behavior of ternary solutions and control the nucleation and removal of nanodroplets. It may be potential in many applications, such as drug delivery, nanoparticle preparation and so on.

\printcredits

\section*{Declaration of Competing Interest}

The authors declare that they have no known competing financial interests or personal relationships that could have appeared to influence the work reported in this paper.

\section*{Acknowledgments}
This work was financially supported by the National Natural Science Foundation of China under Grant Nos. 11988102 and 12202244, and the New Cornerstone Science Foundation through the XPLORER PRIZE.  



\bibliographystyle{elsarticle-num}



\begin{thebibliography}{10}
\expandafter\ifx\csname url\endcsname\relax
  \def\url#1{\texttt{#1}}\fi
\expandafter\ifx\csname urlprefix\endcsname\relax\def\urlprefix{URL }\fi
\expandafter\ifx\csname href\endcsname\relax
  \def\href#1#2{#2} \def\path#1{#1}\fi

\bibitem{bouchemal2004nano}
K.~Bouchemal, S.~Brian{\c{c}}on, E.~Perrier, H.~Fessi, Nano-emulsion
  formulation using spontaneous emulsification: solvent, oil and surfactant
  optimisation, International journal of pharmaceutics 280~(1-2) (2004)
  241--251.

\bibitem{srivastava2006detergency}
V.~K. Srivastava, G.~Kini, D.~Rout, Detergency in spontaneously formed
  emulsions, Journal of colloid and interface science 304~(1) (2006) 214--221.

\bibitem{anton2008design}
N.~Anton, J.-P. Benoit, P.~Saulnier, Design and production of nanoparticles
  formulated from nano-emulsion templates—a review, Journal of controlled
  release 128~(3) (2008) 185--199.

\bibitem{anton2009universality}
N.~Anton, T.~F. Vandamme, The universality of low-energy nano-emulsification,
  International journal of pharmaceutics 377~(1-2) (2009) 142--147.

\bibitem{xu2017synthesis}
J.~Xu, H.~Deng, J.~Song, J.~Zhao, L.~Zhang, W.~Hou, Synthesis of hierarchical
  flower-like mg2al-cl layered double hydroxide in a surfactant-free reverse
  microemulsion, Journal of colloid and interface science 505 (2017) 816--823.

\bibitem{liu2018co2}
D.~Liu, Z.~Huang, Y.~Suo, P.~Zhu, J.~Tan, H.~Lu, Co2-responsive surfactant-free
  microemulsion, Langmuir 34~(30) (2018) 8910--8916.

\bibitem{yan2021nanoprecipitation}
X.~Yan, J.~Bernard, F.~Ganachaud, Nanoprecipitation as a simple and
  straightforward process to create complex polymeric colloidal morphologies,
  Advances in Colloid and Interface Science 294 (2021) 102474.

\bibitem{hoar1943transparent}
T.~Hoar, J.~Schulman, Transparent water-in-oil dispersions: the oleopathic
  hydro-micelle, Nature 152~(3847) (1943) 102--103.

\bibitem{moulik1998structure}
S.~P. Moulik, B.~K. Paul, Structure, dynamics and transport properties of
  microemulsions, Advances in Colloid and Interface science 78~(2) (1998)
  99--195.

\bibitem{subramanian2013mesoscale}
D.~Subramanian, C.~T. Boughter, J.~B. Klauda, B.~Hammouda, M.~A. Anisimov,
  Mesoscale inhomogeneities in aqueous solutions of small amphiphilic
  molecules, Faraday discussions 167 (2013) 217--238.

\bibitem{schoettl2014emergence}
S.~Schoettl, J.~Marcus, O.~Diat, D.~Touraud, W.~Kunz, T.~Zemb, D.~Horinek,
  Emergence of surfactant-free micelles from ternary solutions, Chemical
  Science 5~(8) (2014) 2949--2954.

\bibitem{hou2016surfactant}
W.~Hou, J.~Xu, Surfactant-free microemulsions, Current Opinion in Colloid \&
  Interface Science 25 (2016) 67--74.

\bibitem{zemb2016explain}
T.~N. Zemb, M.~Klossek, T.~Lopian, J.~Marcus, S.~Sch{\"o}ettl, D.~Horinek,
  S.~F. Prevost, D.~Touraud, O.~Diat, S.~Mar{\v{c}}elja, et~al., How to explain
  microemulsions formed by solvent mixtures without conventional surfactants,
  Proceedings of the National Academy of Sciences 113~(16) (2016) 4260--4265.

\bibitem{rak2018mesoscale}
D.~Rak, M.~Sedlak, On the mesoscale solubility in liquid solutions and
  mixtures, The Journal of Physical Chemistry B 123~(6) (2018) 1365--1374.

\bibitem{hodgdon2007hydrotropic}
T.~K. Hodgdon, E.~W. Kaler, Hydrotropic solutions, Current opinion in colloid
  \& interface science 12~(3) (2007) 121--128.

\bibitem{xu2013surfactant}
J.~Xu, A.~Yin, J.~Zhao, D.~Li, W.~Hou, Surfactant-free microemulsion composed
  of oleic acid, n-propanol, and h2o, The Journal of Physical Chemistry B
  117~(1) (2013) 450--456.

\bibitem{eastoe2011action}
J.~Eastoe, M.~H. Hatzopoulos, P.~J. Dowding, Action of hydrotropes and
  alkyl-hydrotropes, Soft Matter 7~(13) (2011) 5917--5925.

\bibitem{kunz2016hydrotropes}
W.~Kunz, K.~Holmberg, T.~Zemb, Hydrotropes, Current Opinion in Colloid \&
  Interface Science 22 (2016) 99--107.

\bibitem{smith1977oil}
G.~D. Smith, C.~E. Donelan, R.~E. Barden, Oil-continuous microemulsions
  composed of hexane, water, and 2-propanol, Journal of Colloid and Interface
  Science 60~(3) (1977) 488--496.

\bibitem{klossek2012structure}
M.~L. Klossek, D.~Touraud, T.~Zemb, W.~Kunz, Structure and solubility in
  surfactant-free microemulsions, ChemPhysChem 13~(18) (2012) 4116--4119.

\bibitem{xu2018surfactant}
J.~Xu, J.~Song, H.~Deng, W.~Hou, Surfactant-free microemulsions of
  1-butyl-3-methylimidazolium hexafluorophosphate, diethylammonium formate, and
  water, Langmuir 34~(26) (2018) 7776--7783.

\bibitem{yan2017modular}
X.~Yan, P.~Alcouffe, G.~Sudre, L.~David, J.~Bernard, F.~Ganachaud, Modular
  construction of single-component polymer nanocapsules through a one-step
  surfactant-free microemulsion templated synthesis, Chemical Communications
  53~(8) (2017) 1401--1404.

\bibitem{novikov2017dual}
A.~A. Novikov, A.~P. Semenov, V.~Monje-Galvan, V.~N. Kuryakov, J.~B. Klauda,
  M.~A. Anisimov, Dual action of hydrotropes at the water/oil interface, The
  Journal of Physical Chemistry C 121~(30) (2017) 16423--16431.

\bibitem{robertson2016mesoscale}
A.~E. Robertson, D.~H. Phan, J.~E. Macaluso, V.~N. Kuryakov, E.~V. Jouravleva,
  C.~E. Bertrand, I.~K. Yudin, M.~A. Anisimov, Mesoscale solubilization and
  critical phenomena in binary and quasi-binary solutions of hydrotropes, Fluid
  Phase Equilibria 407 (2016) 243--254.

\bibitem{prevost2016small}
S.~Prevost, T.~Lopian, M.~Pleines, O.~Diat, T.~Zemb, Small-angle scattering and
  morphologies of ultra-flexible microemulsions, Journal of applied
  crystallography 49~(6) (2016) 2063--2072.

\bibitem{diat2013octanol}
O.~Diat, M.~L. Klossek, D.~Touraud, B.~Deme, I.~Grillo, W.~Kunz, T.~Zemb,
  Octanol-rich and water-rich domains in dynamic equilibrium in the pre-ouzo
  region of ternary systems containing a hydrotrope, Journal of applied
  Crystallography 46~(6) (2013) 1665--1669.

\bibitem{subramanian2014mesoscale}
D.~Subramanian, J.~B. Klauda, P.~J. Collings, M.~A. Anisimov, Mesoscale
  phenomena in ternary solutions of tertiary butyl alcohol, water, and
  propylene oxide, The Journal of Physical Chemistry B 118~(22) (2014)
  5994--6006.

\bibitem{han2022formation}
Y.~Han, N.~Pan, D.~Li, S.~Liu, B.~Sun, J.~Chai, D.~Li, Formation mechanism of
  surfactant-free microemulsion and a judgment on whether it can be formed in
  one ternary system, Chemical Engineering Journal 437 (2022) 135385.

\bibitem{schottl2016aggregation}
S.~Sch{\"o}ttl, D.~Horinek, Aggregation in detergent-free ternary mixtures with
  microemulsion-like properties, Current Opinion in Colloid \& Interface
  Science 22 (2016) 8--13.

\bibitem{jungwirth2006specific}
P.~Jungwirth, D.~J. Tobias, Specific ion effects at the air/water interface,
  Chemical reviews 106~(4) (2006) 1259--1281.

\bibitem{marcelja1997hydration}
S.~Marcelja, Hydration in electrical double layers, Nature 385~(6618) (1997)
  689--690.

\bibitem{marvcelja2011hydration}
S.~Mar{\v{c}}elja, Hydration forces near charged interfaces in terms of
  effective ion potentials, Current opinion in colloid \& interface science
  16~(6) (2011) 579--583.

\bibitem{donaldson2015developing}
S.~H. Donaldson~Jr, A.~R{\o}yne, K.~Kristiansen, M.~V. Rapp, S.~Das, M.~A.
  Gebbie, D.~W. Lee, P.~Stock, M.~Valtiner, J.~Israelachvili, Developing a
  general interaction potential for hydrophobic and hydrophilic interactions,
  Langmuir 31~(7) (2015) 2051--2064.

\bibitem{lopian2016morphologies}
T.~Lopian, S.~Schottl, S.~Prevost, S.~Pellet-Rostaing, D.~Horinek, W.~Kunz,
  T.~Zemb, Morphologies observed in ultraflexible microemulsions with and
  without the presence of a strong acid, ACS central science 2~(7) (2016)
  467--475.

\bibitem{qiao2015molecular}
B.~Qiao, G.~Ferru, M.~Olvera de~la Cruz, R.~J. Ellis, Molecular origins of
  mesoscale ordering in a metalloamphiphile phase, ACS Central Science 1~(9)
  (2015) 493--503.

\bibitem{sweatman2014cluster}
M.~B. Sweatman, R.~Fartaria, L.~Lue, Cluster formation in fluids with competing
  short-range and long-range interactions, The Journal of chemical physics
  140~(12) (2014) 03B626\_1.

\bibitem{sweatman2019giant}
M.~B. Sweatman, L.~Lue, The giant salr cluster fluid: a review, Advanced Theory
  and Simulations 2~(7) (2019) 1900025.

\bibitem{gebauer2008stable}
D.~Gebauer, A.~Volkel, H.~Colfen, Stable prenucleation calcium carbonate
  clusters, Science 322~(5909) (2008) 1819--1822.

\bibitem{davey2013nucleation}
R.~J. Davey, S.~L. Schroeder, J.~H. Ter~Horst, Nucleation of organic
  crystals—a molecular perspective, Angewandte Chemie International Edition
  52~(8) (2013) 2166--2179.

\bibitem{jawor2015effect}
A.~Jawor-Baczynska, B.~D. Moore, J.~Sefcik, Effect of mixing, concentration and
  temperature on the formation of mesostructured solutions and their role in
  the nucleation of dl-valine crystals, Faraday Discussions 179 (2015)
  141--154.

\bibitem{li2022spontaneously}
M.~Li, L.~Yi, C.~Sun, Spontaneously formed multiscale nano-domains in
  monophasic region of ternary solution, Journal of Colloid and Interface
  Science 628 (2022) 223--235.

\bibitem{lohse2020physicochemical}
D.~Lohse, X.~Zhang, Physicochemical hydrodynamics of droplets out of
  equilibrium, Nature Reviews Physics 2~(8) (2020) 426--443.

\bibitem{siedentopf1902uber}
H.~Siedentopf, R.~Zsigmondy, Uber sichtbarmachung und gr{\"o}{\ss}enbestimmung
  ultramikoskopischer teilchen, mit besonderer anwendung auf
  goldrubingl{\"a}ser, Annalen der Physik 315~(1) (1902) 1--39.

\bibitem{schindelin2012fiji}
J.~Schindelin, I.~Arganda-Carreras, E.~Frise, V.~Kaynig, M.~Longair,
  T.~Pietzsch, S.~Preibisch, C.~Rueden, S.~Saalfeld, B.~Schmid, et~al., Fiji:
  an open-source platform for biological-image analysis, Nature methods 9~(7)
  (2012) 676--682.
\newblock \href {https://doi.org/10.1038/nmeth.2019}
  {\path{doi:10.1038/nmeth.2019}}.

\bibitem{wagner2014dark}
T.~Wagner, H.-G. Lipinski, M.~Wiemann, Dark field nanoparticle tracking
  analysis for size characterization of plasmonic and non-plasmonic particles,
  Journal of nanoparticle research 16~(5) (2014) 2419.
\newblock \href {https://doi.org/10.1007/s11051-014-2419-x}
  {\path{doi:10.1007/s11051-014-2419-x}}.

\bibitem{lohse2015surface}
D.~Lohse, X.~Zhang, et~al., Surface nanobubbles and nanodroplets, Reviews of
  modern physics 87~(3) (2015) 981.

\bibitem{kerschbaum2004measurement}
S.~Kerschbaum, G.~Rinke, Measurement of the temperature dependent viscosity of
  biodiesel fuels, Fuel 83~(3) (2004) 287--291.

\bibitem{restolho2009viscosity}
J.~Restolho, A.~P. Serro, J.~L. Mata, B.~Saramago, Viscosity and surface
  tension of 1-ethanol-3-methylimidazolium tetrafluoroborate and
  1-methyl-3-octylimidazolium tetrafluoroborate over a wide temperature range,
  Journal of Chemical \& Engineering Data 54~(3) (2009) 950--955.

\bibitem{yuan2009predicting}
W.~Yuan, A.~Hansen, Q.~Zhang, Predicting the temperature dependent viscosity of
  biodiesel fuels, Fuel 88~(6) (2009) 1120--1126.

\bibitem{hsu2022observation}
W.-H. Hsu, T.-C. Yen, C.-C. Chen, C.-W. Yang, C.-K. Fang, S.~Hwang, Observation
  of mesoscopic clathrate structures in ethanol-water mixtures, Journal of
  Molecular Liquids 366 (2022) 120299.

\bibitem{ma2022measurement}
X.-t. Ma, M.-b. Li, C.~Sun, Measurement and characterization of bulk
  nanobubbles by nanoparticle tracking analysis method, Journal of
  Hydrodynamics (2022) 1--13.

\bibitem{chiappisi2018looking}
L.~Chiappisi, I.~Grillo, Looking into limoncello: the structure of the italian
  liquor revealed by small-angle neutron scattering, ACS omega 3~(11) (2018)
  15407--15415.

\bibitem{schottl2019combined}
S.~Sch{\"o}ttl, T.~Lopian, S.~Pr{\'e}vost, D.~Touraud, I.~Grillo, O.~Diat,
  T.~Zemb, D.~Horinek, Combined molecular dynamics (md) and small angle
  scattering (sas) analysis of organization on a nanometer-scale in ternary
  solvent solutions containing a hydrotrope, Journal of colloid and interface
  science 540 (2019) 623--633.

\bibitem{prevost2021spontaneous}
S.~Pr{\'e}vost, S.~Krickl, S.~Marcelja, W.~Kunz, T.~Zemb, I.~Grillo,
  Spontaneous ouzo emulsions coexist with pre-ouzo ultraflexible
  microemulsions, Langmuir 37~(13) (2021) 3817--3827.

\bibitem{hahn2019ab}
M.~Hahn, S.~Krickl, T.~Buchecker, G.~Jo{\v{s}}t, D.~Touraud, P.~Bauduin,
  A.~Pfitzner, A.~Klamt, W.~Kunz, Ab initio prediction of structuring/mesoscale
  inhomogeneities in surfactant-free microemulsions and hydrogen-bonding-free
  microemulsions, Physical Chemistry Chemical Physics 21~(15) (2019)
  8054--8066.

\bibitem{frenkel1939general}
J.~Frenkel, A general theory of heterophase fluctuations and pretransition
  phenomena, The Journal of Chemical Physics 7~(7) (1939) 538--547.

\bibitem{gebauer2014pre}
D.~Gebauer, M.~Kellermeier, J.~D. Gale, L.~Bergstr{\"o}m, H.~C{\"o}lfen,
  Pre-nucleation clusters as solute precursors in crystallisation, Chemical
  Society Reviews 43~(7) (2014) 2348--2371.

\end{thebibliography}

\end{sloppypar}

\end{document}